# Comprehensive Landscapes for Software-related Quality Models

Michael Kläs, Jens Heidrich, Jürgen Münch, Adam Trendowicz

Fraunhofer Institute for Experimental Software Engineering,
Fraunhofer-Platz 1, 67663 Kaiserslautern, Germany
{michael.klaes, juergen.muench, jens.heidrich, adam.trendowicz}@iese.fraunhofer.de

**Abstract.** Managing quality (such as service availability or process adherence) during the development, operation, and maintenance of software(-intensive) systems and services is a challenging task. Although many organizations need to define, control, measure, and improve various quality aspects of their development artifacts and processes, nearly no guidance is available on how to select, adapt, define, combine, use, and evolve quality models. Catalogs of quality models as well as selection and tailoring processes are widely missing. One essential reason for this tremendous lack of support is that software development is a highly context-dependent process. Therefore, quality models always need to be adaptable to the respective project goals and contexts. A first step towards better support for selecting and adapting quality models can be seen in a classification of existing quality models, especially with respect to their suitability for different purposes and contexts. Such a classification of quality models can be applied to provide an integrated overview of the variety of quality models. This article presents the idea of so called comprehensive quality model landscapes (CQMLs), which provide a classification scheme for quality models and help to get an overview of existing quality models and their relationships. The article describes the usage goals for such landscapes, presents a classification scheme, presents the initial concept of such landscapes, illustrates the concept with selected examples, and sketches open questions and future work.

## 1 Introduction

The multitude of software-related quality models available and the lack of guidance for identifying, evaluating, selecting, and adapting a set of appropriate models for a specific organization or project implies 1) a need for getting a structured classification and overview of available quality models, 2) a need for linking quality aspects to higher-level goals of a project or an organization and the respective context, and 3) a need for appropriate selection and customization processes. This article focuses on the need for getting a structured classification and overview of available quality models.

Currently, a variety of quality models exists, originating from the literature, company standards, official standards, or other sources (e.g., they might be implicitly defined in measurement systems, key performance indicators, or quality gates). Typically, quality models focus on product quality (e.g., [1], [2]), process quality (e.g., maturity models, process adherence, or performance models), resource quality (e.g., server availability model, qualification model), or project quality (e.g., milestone slippage model or project cost). Each of these models usually supports only a limited set of application purposes (like characterization [3], improvement [4], or prediction [5]). In many cases, it is not obvious for which usage purposes the models are suitable, in which contexts they can be applied (e.g., in which application domain), and how to customize them. In addition, it is not clear to what extent the models have already been evaluated. In case of the availability of empirical evidence, evaluation and dissemination are typically limited to a specific context and difficult to find in the literature.

In consequence, quality assurance managers, quality managers, and project planners have significant problems in identifying the appropriate set of quality models that is relevant for them. Furthermore, the lack of a uniform classification of quality models aggravates the communication regarding quality aspects.

The concept sketched in this article consists of a classification of quality models and the use of this classification in a so-called comprehensive quality model landscape (CQML). The idea of cartographing quality models in landscapes can be compared with so-called Enterprise IT landscapes that aim at improving the communication regarding networked IT systems in a company through graphical visualizations [6].

The article is structured as follows: Section 2 defines a set of usage scenarios for CQMLs, Section 3 sketches the classification scheme, Section 4 gives an overview of the landscape concept, and Section 5 provides conclusions, open questions, and directions for future work.

## 2    Usage Scenarios for Comprehensive QM Landscapes

In the context of software engineering, a number of potential decision-making processes may be effectively supported by QM landscapes. In their research, the authors aim at the following particular CQMLs usage scenarios:

**S1**  Improve the communication between involved people (such as quality assurance personnel, managers, project planners, developers, contractors) regarding quality models;
**S2**  Provide a comprehensive overview of (relevant) quality models;
**S3**  Support the identification of gaps (i.e., identify areas where quality models would be necessary but are currently missing);
**S4**  Support the identification and selection of relevant quality models in a goal oriented way;
**S5**  Support the adaptation of quality models to specific goals and contexts (this also requires, for instance, an adaptation process);
**S6**  Support the identification of relationships between quality models;

**S7** Support the combination of quality models (this also requires, for instance, aggregation and composition models as well as a kind of unique format for describing quality models).

## 3 Classification Scheme for Quality Models

The authors see the following initial requirements for the classification and the landscape: Classification categories need to be (1) meaningful/minimal in the sense that they contribute to at least one usage scenario, (2) complete in the sense that all quality models can be categorized, and (3) orthogonal in the sense that the classification is as unambiguous as possible.

In order to systematize and evaluate quality models, we created a classification scheme including major dimensions based on the goal template provided by the well-established Goal-Question-Metric (GQM) paradigm [7]. The GQM goal template specifies five aspects that should be considered while defining goals of software measurement. We utilize the GQM goal template as follows:

- *Object* specifies what is being examined by a quality model. The major classes of objects include products, processes, resources, and projects.
- *Purpose* specifies the intent/motivation of quality modeling. The (initial) major purposes include (ordered by an organization's measuring capabilities):
  - Characterization - describing objects with respect to quality,
  - Understanding - explaining dependencies between (sub-)qualities of objects,
  - Evaluation - assessing the achievement of quality goals,
  - Prediction - estimating the expected value of quality,
  - Improvement - determining what needs to be done for improving quality (quantitatively).
- *Quality Focus* specifies the quality attribute being modeled. Example software-related qualities are reliability of products, maturity of processes, productivity of personnel, or cost of projects.
- *Viewpoint (Stakeholder)* specifies the perspective from which the quality attribute is modeled. Typically, the perspective refers to a stakeholder from whose viewpoint the quality attribute is perceived.
- *Context* specifies the environment in which the quality modeling takes place. The context characteristics should, in particular, cover aspects such as:
  - *Scope*, which specifies the comprehension of an organizational and process area covered by a quality model. An example organizational scope could be the whole organization, business unit, group, or project, whereas an example process scope could be process, activity, or task.
  - *Domain*, which specifies the domain(s) a quality model covers (and is intended for). Typical software application domains include: embedded software systems, management & information systems, and web application.

These characteristics serve as a basis for pre-selecting major groups of quality models. As, in practice, each group will probably contain many, largely heterogeneous, models, additional aspects need to be considered in order to select a narrower group of quality models that will fit particular demands and capabilities. It must be ensured that a model's *critical prerequisites* are fulfilled before it can be utilized. One essential aspect to consider are inputs required by a specific quality model. This includes *type* of data (e.g., objective-subjective, categorical-numerical, or certain-uncertain), *amount of data*, and *quality of data* (e.g., completeness). Further, motivated by industrial demands, we propose considering such aspects as empirical evidence and utilization support. *Empirical evidence* specifies the amount and quality of empirical studies proving that a quality model works in practice as expected. *Utilization support* mainly refers to the amount and quality of documentation for a quality model; it may also include the existence of tools supporting the utilization of a quality model.

Finally, dependent on the values of the major characteristics, further aspects may need to be considered. For example, given that a quality model does not perfectly fit particular needs and capabilities its flexibility might be an essential characteristic to consider. *Flexibility* here refers to the extent to which the model (1) is already predefined (i.e., if it represents a fixed-model or a define-your-model approach) and (2) can be adapted to the specific needs and capabilities. Moreover, dependent on the purpose of modeling, the model's *availability* might play an important role. In case of prediction purposes, the earlier a prediction model can be applied (can provide estimates), the better.

## 4   Landscaping Quality Models

As mentioned in Section 1, there exists a variety of different quality models for different application scenarios, and it is a challenging task to create a landscape of existing models. For example, one usage of CQMLs should be the selection of appropriate quality models in a goal-oriented way for measurement goals defined on the basis of GQM (S4). Depending on the specific application scenarios of an organization, different dimensions may be important for creating such a landscape. Furthermore, it can be helpful to restrict the quality models presented in a landscape by fixing one or more dimensions. For instance, one might consider product quality models only (by restricting the "object" dimension) or one might only consider models relevant in a certain context (e.g., by restricting the scope/domain). Quality models may consist of several sub-models (e.g., ISO 9126 has models for internal quality, external quality, and quality in use). Depending on the level of detail of a particular landscape, those sub-models may have to be classified separately.

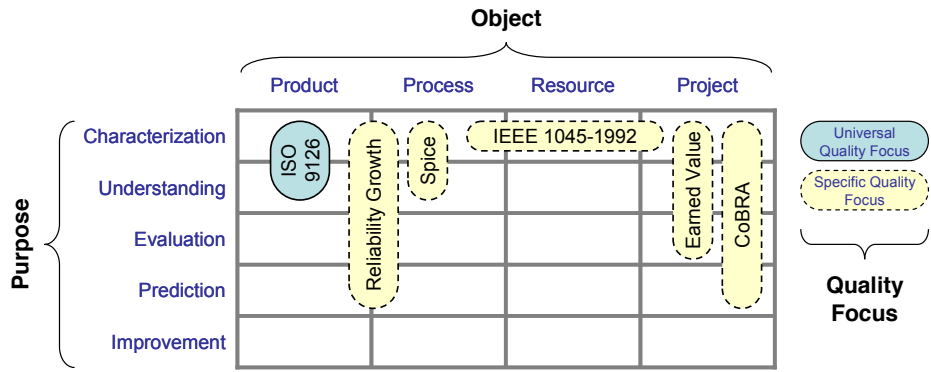

**Fig. 1.** Sample Landscape of Quality Models.

To give an example, let us consider three main dimensions for creating a high-level landscape of quality models: object, purpose, and quality focus. Fig. 1 gives a preliminary classification of some popular quality models according to those three dimensions. The landscape uses quite simple classes for the three dimensions:

- *Object:* A quality model may consider different objects. For instance, ISO 9126 [2] makes statements about the general software product quality, CoBRA [8] addresses project costs, and reliability growth models [5] make statements about actual product reliability or the remaining system test time of the test process. For our landscape, we want to classify the models into four classes: product, process, resource, and project. A quality model may be assigned to more than one class or even only address parts of a class (e.g., a certain sub-process like testing or a certain part of a product like the design document).
- *Purpose:* A quality model may support different purposes. In our landscape, we use the high-level classes: characterization, understanding, evaluation, prediction, and improvement. In practice, a quality model may support more specific purposes (like risk assessment or supplier management) that may be assigned to one or more of the high-level classes presented here. For instance, ISO 9126 may be used to characterize and understand software quality. There are no thresholds defined for the metrics in the model. So, evaluation (comparison) using ISO 9126 is not possible out-of-the-box. Furthermore, no prediction and quantitative improvement scenarios are supported.
- *Quality Focus:* Creating universal classes for the quality focus would probably be as difficult as creating a universal quality model. There exists a variety of qualities and sub-qualities, which are mostly ordered and defined differently in different quality models. For our landscape, we simply want to distinguish between quality models addressing a single, specific quality focus (like productivity as one aspect of process quality [3] or reliability as one aspect of product quality [5]) and models claiming to have a universal quality focus (usually refining qualities by sub-qualities), like the ISO 9126 for product quality.

## 5      Conclusions

This paper presents the first step towards developing a more comprehensive classification scheme for quality models and goal-oriented landscapes (so-called CQMLs). Future work will be in the area of refining and empirically evaluating the classification scheme for accessing and improving the applicability and usefulness for the stated scenarios. Existing quality models should be reviewed and classified based on the validated schema to build up a database of existing quality models. The classification scheme and landscapes should be integrated into a comprehensive model selection and adaptation process (inspired by the comprehensive reuse process [9]) for building up an experience base of quality models. This work is planned to be conducted as part of the publicly funded BMBF project "QuaMoCo".

## 6      Acknowledgments

Parts of this work have been funded by the BMBF project "QuaMoCo – Software-Qualität: Flexible Modellierung und integriertes Controlling" (grant 01 IS 08 023 C).